\documentclass[aps, prx, twocolumn,show pacs,show keys,superscript address,superscript reference,floatfix]{revtex4-2}
\usepackage[colorlinks=true,citecolor=blue,linkcolor=blue,urlcolor=blue]{hyperref}
\usepackage{graphicx,times}
\usepackage[ruled,vlined]{algorithm2e}
\usepackage{color}
\usepackage{amssymb}
\usepackage{tikz,xcolor,hyperref}
\usepackage{subeqnarray}
\usepackage{float}
\usepackage{bbm}
\usepackage{bm}
\usepackage{diagbox}
\usepackage{makecell}
\usepackage[T1]{fontenc}
\usepackage{selinput}
\usepackage{babel}
\usepackage{amsmath}
\usepackage{tabularx, booktabs,lipsum}

\newcommand{\etal}{{\it{et al.}}}
\DeclareMathAlphabet\mathbfcal{OMS}{cmsy}{b}{n}

\definecolor{lime}{HTML}{A6CE39}
\DeclareRobustCommand{\orcidicon}{%
	\begin{tikzpicture}
	\draw[lime, fill=lime] (0,0) 
	circle [radius=0.16] 
	node[white] {{\fontfamily{qag}\selectfont \tiny ID}};
	\draw[white, fill=white] (-0.0625,0.095) 
	circle [radius=0.007];
	\end{tikzpicture}
	\hspace{-2mm}
}

\foreach \x in {A, ..., Z}{%
	\expandafter\xdef\csname orcid\x\endcsname{\noexpand\href{https://orcid.org/\csname orcidauthor\x\endcsname}{\noexpand\orcidicon}}
}

\begin{document}
\title{Digital-Analog Counterdiabatic Quantum Optimization with Trapped Ions}
\author{Shubham~Kumar \orcidA{}}
\affiliation{Kipu Quantum, Greifswalderstrasse 226, 10405 Berlin, Germany}

\author{Narendra N. Hegade \orcidB{}}
\email{narendrahegade5@gmail.com}
\affiliation{Kipu Quantum, Greifswalderstrasse 226, 10405 Berlin, Germany}

\author{Alejandro Gomez Cadavid \orcidC{}}
\affiliation{Kipu Quantum, Greifswalderstrasse 226, 10405 Berlin, Germany}
\affiliation{Department of Physics, University of the Basque Country UPV/EHU, Barrio Sarriena, s/n, 48940 Leioa, Biscay, Spain}

\author{Murilo Henrique de Oliveira \orcidD{}}
\affiliation{Kipu Quantum, Greifswalderstrasse 226, 10405 Berlin, Germany}

\author{Enrique Solano \orcidF{}}
\affiliation{Kipu Quantum, Greifswalderstrasse 226, 10405 Berlin, Germany}

\author{F.~Albarr\'an-Arriagada \orcidE{}}
\affiliation{Departamento de F\'isica, CEDENNA, Universidad de Santiago de Chile (USACH), Avenida V\'ictor Jara 3493, 9170124, Santiago, Chile}

\begin{abstract}

We introduce a hardware-specific, problem-dependent digital-analog quantum algorithm of a counterdiabatic quantum dynamics tailored for optimization problems. Specifically, we focus on trapped-ion architectures, taking advantage from global Mølmer-Sørensen gates as the analog interactions complemented by digital gates, both of which are available in the state-of-the-art technologies. We show an optimal configuration of analog blocks and digital steps leading to a substantial reduction in circuit depth compared to the purely digital approach. This implies that, using the proposed encoding, we can address larger optimization problem instances, requiring more qubits, while preserving the coherence time of current devices. Furthermore, we study the minimum gate fidelity required by the analog blocks to outperform the purely digital simulation, finding that it is below the best fidelity reported in the literature. To validate the performance of the digital-analog encoding, we tackle the maximum independent set problem, showing that it requires fewer resources compared to the digital case. This hybrid co-design approach paves the way towards quantum advantage for efficient solutions of quantum optimization problems.

\end{abstract}

\maketitle
\section{Introduction}
Optimization problems play a central role in several scientific, engineering, and economic applications. Among the vast list of optimization problems, the quadratic unconstrained binary optimization (QUBO) formulation is particularly relevant, not only because of the number of industrial applications that can be mapped to it~\cite{Lucas.2014}, but also because most of the existing quantum hardware can encode such problems \cite{Amin.2023, Henriet.2020,Gomez2023arXiv,Guan2023arXiv}. Quantum computing aims to address this class of optimization problems, where one can take advantage of the all-to-all connectivity presented by trapped-ion quantum processors \cite{Moll.2018, Wiebe.2012, Montanaro.2016, Jens.2024}. 

In the last years, the use of control techniques to enhance quantum algorithms for optimization problems has emerged. Specifically, we highlight the digitized counterdiabatic quantum computing (DCQC) paradigm~\cite{Naren.2021}, where polynomial scaling enhancement has been reported in comparison to conventional methods~\cite{Naren.2022}. DCQC has been used in several contexts from protein folding to factorization, showing the potential and flexibility of the technique in current digital quantum computers \cite{Pranav.2023, DCQF.2023}.

However, with the current noisy intermediate-scale quantum (NISQ) computers, implementing scalable digital methods is not feasible within the available coherence times due to the substantial circuit depth required. Quantum algorithms using just single-qubit and two-qubit gates cause an overhead in the number of gates restricting the size of the problems that can be addressed on current quantum hardware. To counteract these limitations, there is a growing interest in leveraging multiqubit gate operations, or analog blocks to reduce the circuit depth by the simultaneous entanglement of multiple qubits through a single operation, offering a more efficient encoding. Complementing analog interactions with digital gates can reduce the circuit depth, and meanwhile address a variety of problems. It means that a suitable analog block reduces the circuit depth, increasing the success of the algorithms, maintaining a reasonable flexibility and allowing to address a variety of problems. 

The digital-analog quantum computing (DAQC) paradigm~\cite{Adrian.2020} was proposed and implemented to enhance quantum algorithms, including the case of quantum simulations~\cite{Pagano.2021, Solano.2016, Jing.2021, Sanz.2023,Ana.2020, Ana.2022, Ana.2023,Tasio.2021, Rajabi.2019, David.2014,Kumar.2023,Elfing.2024}. Along these lines, DAQC was able to show the potential, flexibility and feasibility offered by current NISQ processors. DAQC solutions require careful design because it is not only hardware dependent, since analog blocks are hardware specific, but also problem dependent, due to specific interactions in the target Hamiltonian.

In this work, we introduce a hardware-specific DAQC protocol to solve larger instances of QUBO problems in trapped ions with the aid of counterdiabatic quantum dynamics. We show the advantage of using Global Mølmer-Sørensen (GMS) gates \cite{Mølmer.1999, Sørensen.1999, Monz.2011, Nam.2018} to obtain many-body interactions which are pivotal in different approximations of the counterdiabatic expansion. We complement it with digital steps in an optimal configuration to achieve a substantial reduction in algorithmic circuit depth. This leads to targeting larger instances of the optimization problem within the available coherence time of the quantum processor. Furthermore, it lays a strong foundation on how to engineer and deploy the digital-analog quantum algorithm on any current NISQ processor, along its specifications, to target larger problem instances. 

\section{Counterdiabatic Drivings}
We focus on QUBO problems whose solution can be encoded in the ground state of the Hamiltonian
\begin{equation}
H_f=\sum_{i < j} J_{ij} \sigma_i^z \sigma_j^z + 
        \sum_i h_i \sigma_i^z .
        \label{Eq.2}
\end{equation}
Here, $J_{ij}$, is the coupling strength between the $i$-th and $j$-th qubit, and $h_i$ is the local energy of the $i$-th qubit, $\sigma_{i}^z$ are the Pauli $Z$ operators for the $i$-th qubit. To reach the ground state, we can perform an adiabatic evolution governed by the time-dependent Hamiltonian
\begin{eqnarray}\label{eq:adiabatic_hamiltonian}
    H_{\text{ad}}(t) = &&\lambda(t) \left(\sum_{i<j} J_{ij} \sigma_i^z \sigma_j^z + \sum_i h_i \sigma_i^z \right)\nonumber \\
    &&+ \left[1 - \lambda(t)\right] \sum_i \sigma_i^x,
\end{eqnarray}
with initial state $\otimes _{j}|-_j\rangle$, where $|-_j\rangle=\frac{1}{\sqrt{2}}(|0_j\rangle - |1_j\rangle)$. The scheduling function $\lambda(t)$ is in principle an arbitrary function with the boundary conditions $\lambda(0)=0$ and $\lambda(T)=1$ with $T$, the total time of the adiabatic evolution.

To accelerate this optimization process, we can use shortcut-to-adiabaticity techniques, for example, to add counterdiabatic (CD) terms~\cite{Naren.2021, Naren.2022, Rice.2003, Berry.2009, Muga.2010, Campo.2013}. That is, to evolve the system under the following Hamiltonian
\begin{equation}
    H(t) = H_{\text{ad}}(t) + \dot{\lambda}(t)A_{\lambda}(t),
    \label{Eq.3}
\end{equation}
instead of the purely adiabatic method. Here, $A_{\lambda}$ is called the adiabatic gauge potential (AGP). There exists several approximations to the AGP such as mean field approximation, variational approach~\cite{Sels2017PNAS}, and nested commutator expansion~\cite{Claeys2019PhysRevLett}. Recently, the last one has gained relevance since it provides an approximation with different complexity in the many-body interaction terms, that is
\begin{equation}
    A_{\lambda}(t)=i\sum_{j}^{\infty}\alpha_j(t)[\underbrace{H_{\text{ad}},[H_{\text{ad}}[...[H_{\text{ad}}}_{2j-1},\partial_{\lambda}H_{\text{ad}}]]]] ,
\end{equation}
where $\alpha_j$ can be determined as proposed in Ref. \cite{Xie2022PhysRevB}. In this approximation, the first order takes the form
\begin{equation}\label{eq:cd_hamiltonian}
    A^{(1)}(t) = 2{\alpha}_1(t) 
    \left[ 
        \sum_i h_i \sigma_i^y 
        + \sum_{i < j} J_{ij} (\sigma_i^y \sigma_j^z + \sigma_i^z \sigma_j^y)
    \right].
\end{equation}
The detailed calculation of the analytical expression of $\alpha_1$ is given in Appendix~\ref{0.A}. Notice that the Hamiltonian of Eq.~(\ref{eq:cd_hamiltonian}) involves only local and bilocal terms, matching with capabilities of current devices. In trapped ions, GMS interactions can realize $\sigma_i^z \sigma_j^z $, $\sigma_i^{z} \sigma_j^{y}$ and $\sigma_i^{y} \sigma_j^{z}$ simultaneously in a single gate operation. Below, we show a digital-analog encoding of the above problem which can be performed in trapped ions.

\begin{figure*}
\centering
\includegraphics[width=1\textwidth]{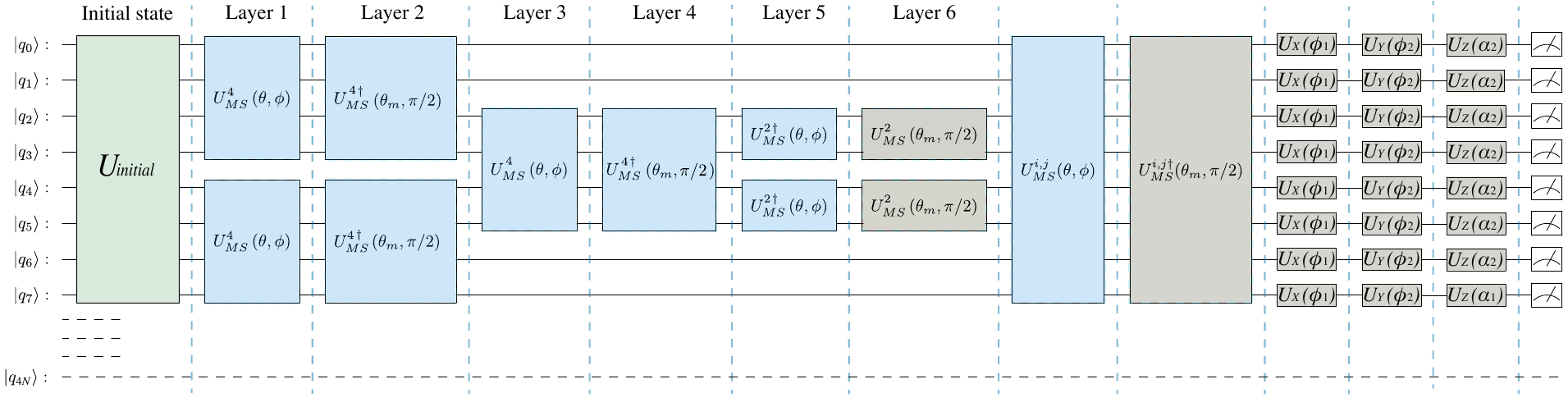}
\caption{Digital-analog quantum algorithm to perform CD-based optimization of homogeneous Ising Hamiltonian in trapped ions, a model widely utilized in quantum simulations of magnetic materials and combinatorial optimization problems. The circuit is modular and extendable, capable of accommodating an arbitrary number of qubits ($N$) to simulate larger systems. Key components of the circuit include analog blocks, highlighted in blue, which are responsible for the execution of continuous-time quantum operations that directly correspond to the $2-local$ interactions in the Ising model and the digital blocks (highlighted in grey) are integrated within a digital framework that provides precise control and measurement capabilities. The modular nature of this design allows for scalable quantum computing applications, from fundamental physics research to practical problem-solving in industry-related topics.}
\label{fig:1}
\end{figure*}

\section{The algorithm}
We introduce the digital-analog counterdiabatic quantum optimization (DACQO) algorithm. As the main analog block of the algorithm, we utilize the GMS gate as a resource, considering trapped-ion quantum processors as the target. This GMS gate can be generated by applying a global entangling operation over a subset of qubits using a uniform microwave or laser field across a set of ions. This approach allows for the simultaneous manipulation of multiple qubits, facilitating the generation of large-scale entangled states~\cite{Kim.2023, Nam.2018}. The GMS gate, complemented with single qubit rotations, has shown to be advantageous in reducing the gate count in quantum circuits for executing quantum algorithms~\cite{Nam.2018}. In this work, we mix the circuit-depth reduction provided by the GMS and the power of digitized counterdiabatic quantum computing to solve binary optimizations, providing a significant step in the path to quantum advantage with digital-analog quantum computing protocols. A GMS gate is given by
\begin{equation}
U_{MS}^k(\theta, \phi) = \exp \left[ -\frac{i \theta}{4} (\cos \phi S_x + \sin \phi S_y)^2 \right],
\label{Eq.5}
\end{equation}
where, $S_{x,y} = \sum_{i=1}^{k} \sigma_i^{x,y}$ and $k$ is the number of ions (qubits) to which the corresponding gate is applied. The GMS gate produces 2-local terms, as well as effective $n$-local term as was shown in Ref.~\cite{Casanova2012PhysRevLett}, which can be use to implement high-order terms in the AGP approximation. To understand how to get effective $n$-local terms, see supplementary sec.~\ref{1.A} and \ref{1.B} for a complete derivation. 

Since the analog block (GMS gate) can only produce terms of the form $\sigma_i^x \sigma_j^x$, $\sigma_i^x \sigma_j^y$ and $\sigma_i^y \sigma_j^y$, we can write the Hamiltonian for our dynamics as

\begin{eqnarray}\label{Eq.6}
H'(\lambda)& = &\lambda(t) \left[ \sum_{i < j =1}^N J_{ij} \sigma_i^x \sigma_j^x + \sum_{i=1}^N h_i \sigma_i^x \right] \nonumber \\
 &&+ (1 - \lambda(t)) \bigg[ \sum_{i=1}^N \sigma_i^z\bigg] + 2\dot{\lambda} {\alpha}_1(t) \bigg[ \sum_{i=1}^N h_i \sigma_i^y \nonumber \\
 && + \sum_{i < j=1}^N J_{ij} (\sigma_i^y \sigma_j^x + \sigma_i^x \sigma_j^y) \bigg].
\end{eqnarray}
Now, all the $2$-local terms in Eq. (\ref{Eq.6}) can be realized by the analog block in Eq. (\ref{Eq.5}) without extra rotations. Next, we will digitized the evolution governed by the Hamiltonian (\ref{Eq.6}) for two cases, first the homogenous case (best case) and then for the full inhomogenous case (worst case).

\subsection{The homogeneous case}
For the homogeneous case,  we consider $J_{ij} = J, h_{i} = h$ in Eq.~(\ref{Eq.6}). We note that, from Eq. (\ref{Eq.5}), we can obtain bilocal terms $\sigma_{i}^a\sigma_{j}^b$, where $a(b)=\{x,y\}$, with strength $\theta^{a,b}$ given by 
\begin{eqnarray}
\theta \cos^2(\phi) /2 &=& \theta^{x,x}, \nonumber\\
\theta \sin^2(\phi) /2 &=& \theta^{y,y}, \nonumber\\
\theta \cos(\phi) \sin(\phi)/4 &=& \theta^{x,y} = \theta^{y,x}.
\end{eqnarray}

 This maps the Hamiltonian coefficients to the angles of the GMS gate given in Eq. (\ref{Eq.5}), obtaining $\theta^{x,x}  = \lambda(t)  J t/n$ and $\theta^{x,y} = 2 J \dot{\lambda}(t) {\alpha}_1(t) t/n$ with $t$ the time and $n$ the number of trotter steps (see supplementary information Sec.~\ref{1.C}). Now, the local terms can be done by single-qubit rotations with angles $\theta^{x} = \lambda(t) h t/n$, $\theta^z = (1-\lambda(t)) t/n$, $\theta^{y} = 2 \dot{\lambda}(t) {\alpha}_1(t) ht/n$. Finally, the GMS gate also produces an extra $\sigma^{y}\sigma^{y}$ terms, that can be eliminated by applying the conjugate GMS gate $ U_{MS}^{k\dagger}(\theta_m, \pi/2)$, where $\theta_m = \theta \sin^2(\phi)$. Considering all these ingredients for the digital-analog proposal, the algorithms is as follow.
 
{\textit{Algorithm description}:}
The algorithm is depicted in Fig.~\ref{fig:1} and described in \textbf{Algorithm \ref{Algo1}}. We consider experimentally feasible GMS gates as analog block, then we restrict it only to the nearest-neighbour ions up to 4 qubits~\cite{Monz.2011}. Despite this, our aim is to entangle as many ions as possible while still achieving a reasonable fidelity.

\begin{algorithm}
    \caption{Digital-analog counterdiabatic quantum optimization for homogeneous problems}\label{Algo1}
    \textbf{Layer 1: Initial entanglement}\;
    \For{each $k$ set of nearest neighbour qubits}{
        Apply $k$-qubit GMS gates $U_{MS}^{k\dagger}(\theta, \phi)$\;
    }
    \textbf{Layer 2: Parasitic term elimination}\;
    \For{each $k$ set of qubits affected in Layer 1}{
        Apply conjugate GMS gates $U_{MS}^{k\dagger}(\theta_m, \pi/2)$\;
    }
    \textbf{Layer 3: Supplementary entanglement}\;
    \For{remaining nearest neighbour $k$ set of qubits}{
        Entangle using $U_{MS}^{k}(\theta, \phi)$\;
    }
    \textbf{Layer 4: Further parasitic term elimination}\;
    For same set of qubits as layer 3, apply $U_{MS}^{k \dagger}(\theta_m, \pi/2)$\;
    \textbf{Layer 5: Overlap correction}\;
    \For{overlapping GMS gates}{
        Apply 2-qubit GMS gates\;
    }
    \textbf{Layer 6: Final parasitic term elimination}\;
    Target the last set of parasitic terms using $U_{MS}^{2 \dagger}(\theta_m, \pi/2)$.\;
    \textbf{Ion pair entanglement}\;
    Entangle remaining ion pairs using 2-qubit GMS gates\;
    Remove the parasitic terms as in layer 2\;
    \textbf{Apply single qubit gates}\;
    \For{all ions}{
        Apply single-qubit rotations\;
    }
    \textbf{Perform unitary time evolution}\;
\end{algorithm}

\begin{figure}
\centering
\includegraphics[width=0.9\columnwidth]{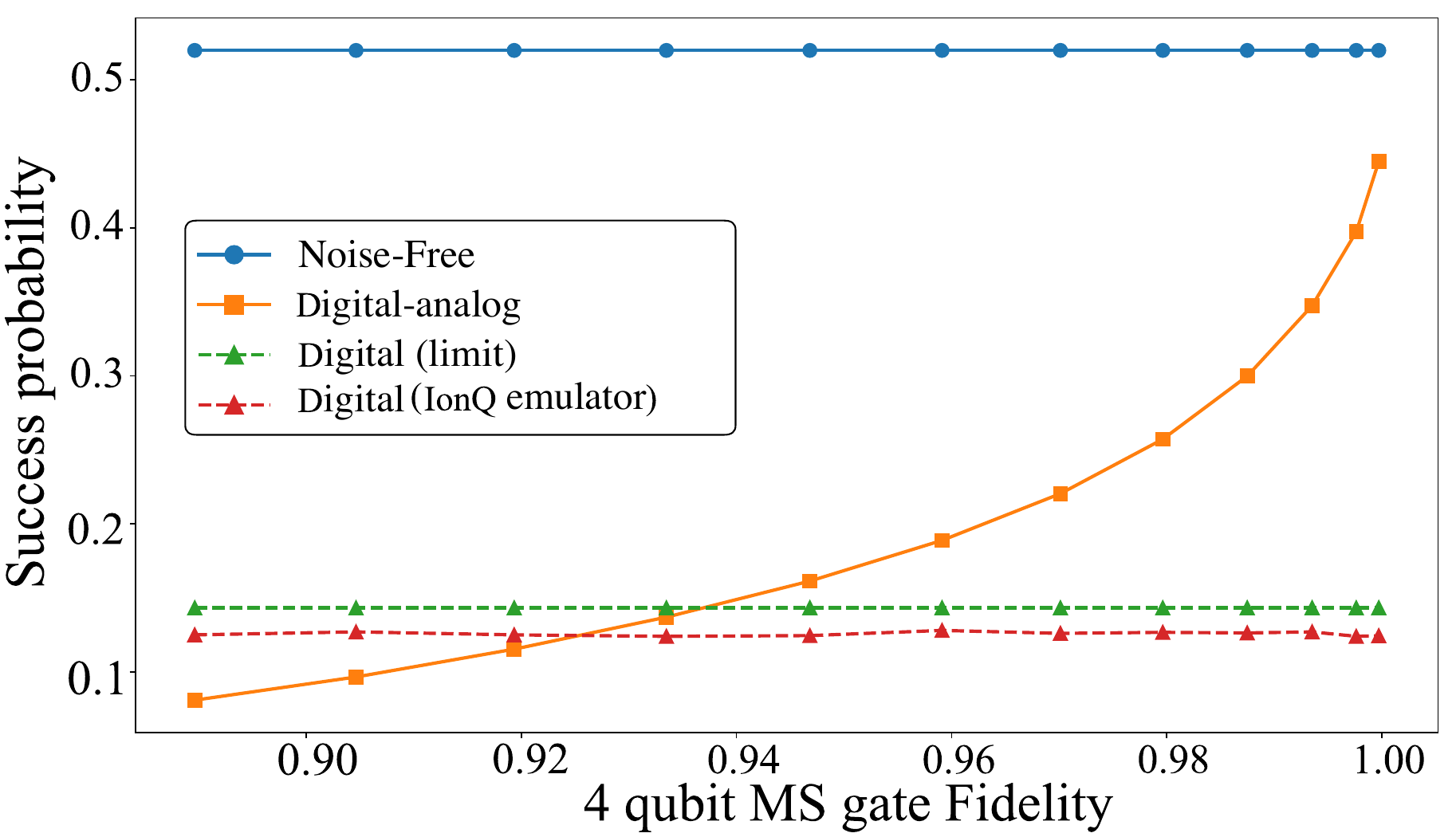}
\caption{DACQO applied to 4 qubit homogeneous Ising model. The figure illustrates the expected success probability as the fidelity of the 4-qubit GMS gate (analog block) improves for 10 trotter step and 1 unit of evolution time. The purpose of the figure is to show the region where digital-analog can surpass purely digital simulation.}
\label{fig:2}
\end{figure}

\begin{figure*}
\centering
\includegraphics[width=1\textwidth]{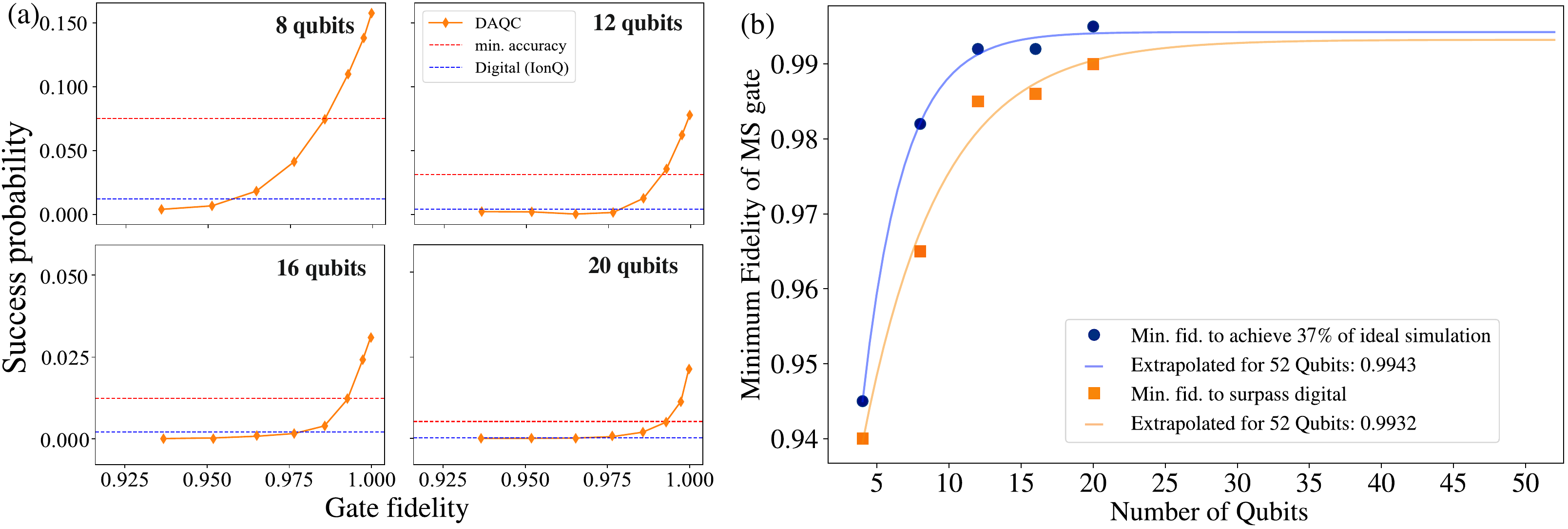}
\caption{Analysis of the success probability and gate Fidelity requirements to run digital-analog counterdiabatic quantum algorithms in trapped ions (2 trotter steps and 15 units of annealing time). (a) This panel illustrates the success probability achievable for varying problem sizes as the GMS gate fidelity increases. The results are compared against the peak performance achievable through purely digital quantum simulations (run on IonQ noisy emulator). Additionally, the figure includes a benchmark indicating the minimum accuracy threshold set at 37 $\%$ of the ideal success probability (current benchmark for circuit success probability by IonQ \cite{IonQ.2024}), highlighting the efficacy of the algorithm in meeting or exceeding this benchmark across different fidelity levels. (b) In this panel, we extrapolate the minimum gate fidelity required for the algorithm to achieve the stated minimum accuracy and to surpass the capabilities of digital simulation, up to a hypothetical 52-qubit system. The extrapolation is based on the observed fidelity trends for problem sizes up to 20 qubits (multiple measurements with 1024 shots and averaged overall), providing insights into the scalability of the algorithm in handling larger quantum systems with stringent accuracy requirements. The extrapolation employs an exponential decay model towards a limit, mathematically represented as $L + (K-L)e^{\lambda N}$, where $L=1$ denotes the limiting fidelity value, $K$ represents the initial fidelity, $\lambda=0.1$ is the decay constant, and $N$ is the number of qubits.} 
\label{fig:3}
\end{figure*}

\textbf{Layer 1} and \textbf{Layer 2} involve the application of the 4-qubit GMS gates. Layer 3 aims to entangle the remaining set of $k$ nearest neighbor qubits as efficiently as possible while minimizing the creation of parasitic terms. \textbf{Layer 5} is designed to eliminate any parasitic terms that arise from the overlap between \textbf{Layer 3} and \textbf{Layer 4}, using conjugate GMS gates.  

After the \textbf{Layer 6}, the number of ion pairs left to be entangled scales as $n_p = {(N-k)(N-k+1)}/2$ and the gate requirement scales as $2n_p$ (including counjugate GMS gates). In trapped ions, ideally $N/2$ gates can be performed simultaneously, this leads to a circuit depth of $\frac{2(N-k)(N-k+1)}{N}$. The total circuit depth for a $N$ qubit system can be expressed as the number of gate layers that can be executed sequentially, where each layer consists of gates that can be executed in parallel as follows
\begin{align}
\text{Depth} &= \text{layers}[U_{MS}^4(\theta)] + \text{layers}[U_{MS}^2(\theta)] + \text{layers}[U_{X,Y,Z}]  \nonumber \\&= 9 + \frac{2(N-k)(N-k+1)}{N}. 
\label{Eq.8}
\end{align}

\begin{algorithm}
    \caption{Digital-analog counterdiabatic quantum optimization for inhomogeneous problems}\label{Algo2}
    \textbf{Step 1: Initial entanglement with inhomogeneity}\;
    \For{each $k$ set of nearest neighbour qubits}{
        Replace homogeneous $U_{MS}^{k\dagger}(\theta, \phi)$ with $k(k-1)/2$ blocks\;
        Apply local rotations to introduce inhomogeneity\;
    }
    \textbf{Step 2: Parasitic term elimination}\;
    \For{each $k$ set of qubits affected in Step 1}{
        Apply conjugate GMS gates $U_{MS}^{k\dagger}(\theta_m, \pi/2)$\;
    }
    \textbf{Step 3: Supplementary entanglement with inhomogeneity}\;
    \For{remaining nearest neighbour $k$ set of qubits}{
        Entangle using $k(k-1)/2$ number of $U_{MS}^{k}(\theta, \phi)$ blocks\;
        Apply local rotations for inhomogeneity\;
    }
    \textbf{Step 4: Further parasitic term elimination}\;
    For same set of qubits as layer 3, apply $U_{MS}^{k \dagger}(\theta_m, \pi/2)$\;
    \textbf{Step 5: Overlap correction}\;
    \For{overlapping GMS gates}{
        Apply 2-qubit GMS gates\;
    }
    \textbf{Step 6: Final parasitic term elimination}\;
    Target the last set of parasitic terms using $U_{MS}^{2 \dagger}(\theta_m, \pi/2)$\;
    \textbf{Ion pair entanglement}\;
    Entangle remaining ion pairs using 2-qubit GMS gates\;
    Remove the parasitic terms as in Step 2\;
    \textbf{Apply single qubit gates}\;
    \For{all ions}{
        Apply single-qubit rotations\;
    }
    \textbf{Perform unitary time evolution}\;
    Apply trotterization to simulate the time evolution\;
\end{algorithm}
As a proof of concept, we solve the 4-qubit QUBO problem using a 4-qubit analog block to demonstrate the efficiency of the algorithm. We perform a noisy simulation by introducing noise using a random matrix noise as $U_{noisy} =c \cdot \mathcal{R}^4$, where $c$ is the noise amplitude and $\mathcal{R}^4$ is a 4-qubit random matrix applied to the GMS gate unitary, Eq. (\ref{Eq.5}). The purpose of this manual noise effect is that it allows us to manually control the noise (tuning $c$) to understand the performance of the algorithm based on the analog gate fidelity, which will give us an approximate idea of the minimal performance required by the analog building blocks to run the algorithm on a hardware. As a realistic noise source, we introduce depolarizing error of magnitude $0.02 \%$ (current state of the art of \textit{IonQ} hardware) local to the qubits (performed using Qiskit's noise model \cite{QiskitAer.2021}). The analysis presented in Fig. \ref{fig:2} illustrates the enhancement of success probability for achieving the ground state as the fidelity of the analog block increases (corresponding to a reduction in noise amplitude). As a benchmark, we consider the digital quantum simulation performed in \textit{IonQ} noisy emulator. This platform provides a $\sigma_i^x\sigma_j^x$ gate with a fidelity ($\mathcal{F}$) of $99.5\%$, and with the corresponding success probability, we can test the performance of the algorithm. Notably, the results demonstrate that DAQC has the potential to surpass digital simulation performance when the analog block fidelities exceed approximately $94\%$. This finding is particularly significant, underscoring the ability to achieve high-accuracy simulations even in the presence of noisy analog blocks, thereby highlighting the robustness and efficiency of digital-analog encoding in leveraging imperfect quantum hardware to realize practical quantum computation. 

We have expanded our analysis to explore the scalability of the algorithm by evaluating the system capacities that can be effectively targeted. To this end, we utilize a 4-qubit analog block to perform DACQO across systems comprising $4N$ qubits (since we use a 4-qubit analog block). This approach enables us to assess the fidelity thresholds necessary for analog blocks to outperform digital simulations and achieve satisfactory accuracy levels. The results of this analysis are presented in Fig.~\ref{fig:3}. Additionally, we have extended our analysis to project the minimum fidelity thresholds necessary for surpassing the performance benchmarks set by digital simulations and meeting the established minimum accuracy standards. These projections, which estimate the fidelity requirements for systems scaling up to 52 qubits, leverage the observed trends in our current data to anticipate the evolving demands on quantum system performance as it expands. The findings indicate that for system sizes up to 20 qubits, GMS gate fidelities within the $98\%-99\%$ range are adequate to surpass purely digital simulations and fidelities between $99\%-99.5\%$ are required to achieve minimum accuracy benchmark. The extrapolated data suggests that to meet or exceed both benchmarks, fidelities around approximately $99.5\%$ are required, which are achievable with current state-of-the-art capabilities~\cite{Bermudez.2017}. This means that for achieving reasonable accuracy, the necessity of engineering very high fidelity 2-qubit gates \cite{Ming.2023} can be relaxed and instead one can focus on engineering analog gates with a reasonable fidelity by using similar techniques, and finally supplement it with the available digital gate sets to perform digital-analog encodings.

{\textit{The inhomogeneous case:}}

Particularly interesting is the case when one would like to tackle an inhomogeneous Hamiltonian, Eq. (\ref{Eq.2}) (different $J_{ij}$'s). The algorithm introduced for the homogeneous case can be extended to the inhomogeneous case. In principle, creating inhomogeneity using a homogeneous analog block will require more gates depending on the size of the analog blocks used. To solve a $N$ qubit inhomogeneous case using a $k-qubit$ analog block, we can create inhomogeneity by applying local rotations to a sequence of analog blocks. 
A simple example to illustrate how to realize the $2-local$ terms for a 3-qubit QUBO Hamiltonian, Eq. (\ref{Eq.6}) using 3-qubit analog blocks is shown in the Supplementary \ref{1.D}. In general, to introduce inhomogeneity across any $k$-qubit interactions within a $N$-qubit system, it is necessary to employ $k(k-1)/2$ analog blocks \cite{Adrian.2020} along with local rotations. Consequently, a pragmatic strategy for addressing inhomogeneous QUBO problems involves the systematic realization of inhomogeneity for each subset of $k$ qubits. This approach is critical for effectively managing the complexity of interactions in any $N$-qubit system. This can be done by replacing only the analog blocks ($U_{MS}^{k\dagger}(\theta, \phi)$) in the \textbf{Algorithm \ref{Algo1}} with $k(k-1)/2$ homogeneous analog blocks supplemented with local rotations. the complete algorithm is explained in \textbf{Algorithm \ref{Algo2}}. The number of local rotations required scales as $k(k-1)$ for each $k$-qubit block. This suggests that bigger analog blocks will incur a large circuit depth. Along these lines, we study the scaling of the DACQO algorithm for homogeneous and inhomogeneous cases and by using different analog block sizes. This will reveal the size of the problem in terms of the number of qubits that can be solved using current trapped ion quantum computers. As a benchmark we will consider the current specifications of $IonQ$ (Forte device).

\begin{figure*}
\centering
\includegraphics[width=1\textwidth]{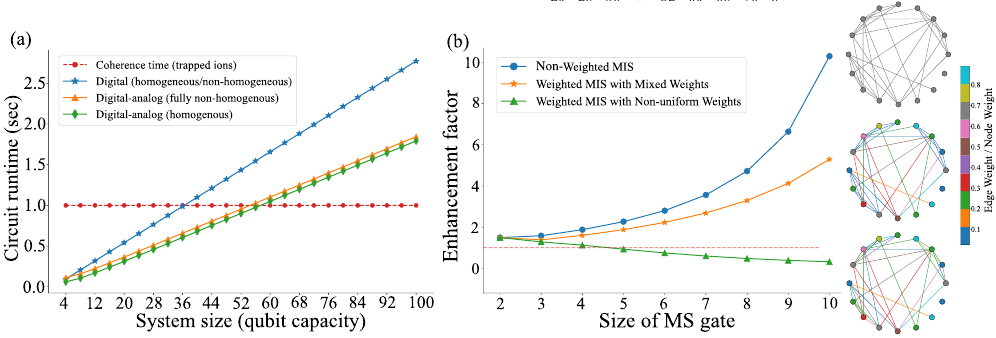}
\caption{Scaling of the algorithm using GMS gates as the analog blocks. (a) Circuit runtime using 4-qubit analog blocks, and (b) enhancement factor for 16 node MIS problem using analog blocks of varying qubit sizes.}
\label{fig:4}
\end{figure*}

\section{Scaling}

We study the scaling of the DACQO algorithm to see the effective system sizes it can solve within the coherence time allowed by current trapped ion hardwares. By accounting for both multi-qubit and single-qubit gate durations, we can calculate the overall circuit runtime in seconds. This quantity can be represented as the sum of the depths multiplied with their duration, with each depth's duration determined by the execution times of respective layers, as shown below:
\begin{eqnarray}
\text{Circuit Runtime (sec)}& = &t_M \cdot \text{depth}[U_{MS}(\theta, \phi)] \nonumber \\ 
&&+ t_S \cdot \text{depth}[U_{X,Y,Z}], 
\end{eqnarray}
where $t_M$ and $t_S$ are the multiqubit and single-qubit gate times, respectively and $layers [U_{MS}(\theta)]$, $t_s (layers[U_{X(Y,Z)}])$ are the number of layers of multi-qubit and single qubit gates. We consider the gate times according to the current state of the art for 2-qubit GMS gates and single qubit gates that are available in current quantum processors (we consider the \textit{IonQ} Forte system for our current analysis). For the 4-qubit GMS gates, we can consider their anticipated gate times achievable with the current technology \cite{Bermudez.2017} which is equivalent to the current 2-qubit gate times. In the current trapped ion systems, for example in the $IonQ$ $Forte$ system, the gate times are $t_M = 930 \mu s$ and $t_S = 130 \mu s$. Based on these numbers achievable with the current technology, we study the circuit runtime of the DACQO algorithm with respect to the size of the optimization problem upto 100 qubits. The circuit runtime plots for both homogeneous and inhomogeneous fully-connected spin glass models are presented in Fig. \ref{fig:4} (a), utilizing 4-qubit GMS gates as the analog blocks. In this figure, the digital-analog encoding, embodying the DACQO paradigm, demonstrates a notably reduced runtime compared to the fully digital paradigm within the permissible coherence time of the current trapped ion processors indicating that employing DAQC can target larger problem instances (upto 55 qubits) requiring more qubits. To target larger instances of the problem, such as those involving 100 qubits, will require extending the processor's coherence time to approximately 1.8 seconds. This represents a significant improvement when compared to the purely digital approach, which requires about 2.8 seconds, thus achieving a $55\%$ reduction in the runtime.

Furthermore, for demonstrating the efficiency of the DACQO algorithm, we consider different block sizes and study the enhancement factor (${\mathcal{R}_{Digital}}/{\mathcal{R}_{DAQC}}$), where $\mathcal{R}$ is the circuit runtime for a fixed problem size. We select the maximum independent set (MIS) problem for this purpose with three different instances involving weighted and non-weighted graphs and 16 nodes. To tackle this problem efficiently, we can make use of the all-to-all connectivity of the trapped-ion quantum processors. We can represent the nodes with equal weights as the nearest neighbour qubits and apply the analog blocks only in $Layer1$, as shown the quantum circuit of Fig. \ref{fig:1}. This is because the problem lacks all-to-all connectivity, necessitating alignment with both the hardware topology and the algorithm's structure, thus exemplifying a co-design approach. The result, shown in Fig. \ref{fig:4} (b) demonstrates that the algorithm shows about $1.6X-$fold improvement in reducing the circuit runtime even with the smallest analog block of 2 qubits. Moreover, further reduction can be achieved with increasing block sizes and the algorithm is most effective for a non-weighted graph followed by a graph with mixed weights and a fully-non uniform graph. For a fully-non uniform graph, increasing the size of the block (beyond 4 qubits) leads to a disadvantage, since the number of gates required for introducing non-homogeniety surpasses the number needed to realize all the many-body terms. Therefore, the algorithm is problem dependent and demonstrates non-trivially that for a fully non-homogeneous problem (non-uniform MIS) the highest reduction can be achieved with just a 2-qubit analog block $[U_{MS}(\theta,\phi)]$.

\section{Future direction}
Although, the hardware-specific implementation of DAQC shows the possibility of solving problems involving qubit capacity beyond 50, it is worthwile to note that this was demonstrated by using a homogeneous multiqubit gate as the analog block applied on the nearest set of qubits. This implies that this predicted improvement is the worst case scenario and, in principle, one can design a programmable inhomogeneous analog block to tackle larger problem instances in trapped ions and the ability to entangle non-nearest neighbour blocks. Recent works on trapped ions have showcased the possibility of fast and programmable ( capable of generating inhomogeneous or specific interactions within a set of ions) multiqubit $XX$ gates~\cite{Ravid.2020, Peleg.2023, Nam.2020, Nam.2022} with an infidelity range of $\sim 10^{-5}-10^{-4}$ upto $4$ ions, which can be used for the further development of DAQC in trapped ions. These gates are also referred to as EASE gates~\cite{Nam.2022} of which the $XX$ type GMS gate is a subset. An estimate of the circuit runtime by employing $4-$qubit programmable $XX$ gates can made. Since it is programmable, both homogeneous and inhomogeneous cases can be tackled with equal number of gates. Using DACQO, where we apply the analog blocks to the nearest neighbour ions, we need three layers of $XX$ gates for each quadrupole of ions (to realize spin-spin and cross Pauli terms). The circuit depth will be
\begin{align}
\text{Depth} &= \text{layers}[XX^4(\theta)] + \text{layers}[U_{MS}^2(\theta)] + \text{layers}[U_{X,Y,Z}]  \nonumber \\&= 6 + \frac{2(N-4)(N-3)}{N}.  
\label{Eq.10}
\end{align}
In this manner, one achieves an improved scaling for the non-homogeneous problems allowing to solve larger problem instances. Furthermore, the ability to generate analog interactions among non-nearest neighbour ions could lead to further reduction depending upon the number of non-nearest pairs that can be entangled. Suppose we apply $4$-qubit analog blocks among non-nearest neighbours, that can generate entanglement among six pairs of ions simultaneously, so that applying $M$ number of such blocks generates $6M$ pairs. Then, the reduced depth will be given by
\begin{align}
\text{Depth} &= 6 + \frac{2[(N-4)(N-3)-6M]}{N}.  
\label{Eq.11}
\end{align}
Thus, the system size that can be targeted within the coherence time of the device will depend on the number $M$. This is related to the ability of the hardware to entangle as many non-nearest neighbour ions as possible based on an analog block of a particular size.

Also, as we mentioned before, GMS gates applied over a set of ions beyond nearest-neighbours. This opens the door to produce effective $k$-local interactions, allowing the implementation of high-order approximations for the AGP, being closer to the exact one, and approaching to the quantum speed limit. Therefore this type of encoding give us a clear path to implement ultrafast algorithms, which are otherwise limited by the technological performance of the GMS interactions.

\section{Conclusions}

We devised a hardware-specific application-dependent quantum algorithm for combinatorial optimization by implementing a digital-analog encoding of the corresponding digitized counterdiabatic quantum optimization algorithm. The aim is to be able to solve larger instances of optimization problems in trapped ions. The encoding is problem dependent, and we tailor the algorithm to solve combinatorial optimization problems such as QUBO. Utilizing the multi-qubit Mølmer-Sørensen interactions in trapped ions as the analog blocks, while supplementing it with single and two-qubit digital steps, we show how to achieve an optimal digital-analog configuration. We design an $N$-qubit quantum circuit and showcase that for QUBO problems. In the worst case scenario, where analog blocks are applied only to nearest-neighbour qubits or ions, we can target larger problem instances requiring more qubits ($\geq 20$) while respecting the available qubit coherence time. We developed the algorithm for a general all-to-all connected Ising spin glass model, and studied the minimum fidelity required by the analog blocks to solve them using the selected hardware. For smaller instances (up to 20 qubits), the algorithm can outperform digital quantum simulation with an average minimum analog block fidelity of $98\%-99\%$ and the extrapolated data (up to 56 qubits) shows that a fidelity of $99\%-99.5\%$ is sufficient. Moreover, we estimate the scaling of an industry relevant use case problem, the maximum independent set by generating a random instance of 16 nodes, and compare its circuit runtime, when the system is 1) non-weighted, 2) weighted with mixed weights, and 3) weighted with non-uniform weights, against the purely digital simulation. In consequence, we demonstrate that the algorithm reduces the circuit runtime by a factor $\sim 2X$ in all instances, and for the first two instances, further reduction can be achieved by employing a larger analog block. In the third instance, however, employing an analog block larger than 4-qubit does not lead to a better scaling. We also suggest potential advancements through the use of fast, programmable inhomogeneous analog blocks \cite{Ravid.2020, Peleg.2023}. 
In summary, our algorithm can reduce the circuit runtime by a minimum factor of $\sim 2X$, enabling the hardware to solve larger problem instances ($\geq 20$ qubits) within the available coherence time.

We believe that the DACQO paradigm paves the way to develop an intensive and interdisciplinary research in digital-analog quantum computing encoding of industry use cases, to bring quantum advantage to the present of academic and commercial quantum processors.

\textit{Acknowledgments.-} F.A.-A. acknowledge the financial support of Agencia Nacional de Investigaci\'on y Desarrollo (ANID): Subvenci\'on a la Instalaci\'on en la Academia SA77210018,  Fondecyt Regular 1231172, Financiamiento Basal para Centros Cient\'ificos y Tecnol\'ogicos de Excelencia AFB220001. We thank useful discussions with Dr. Archismita Dalal, Dr. Anton Simen Albino and Dr. Qi Zhang from Kipu Quantum for useful discussions during the development of this work.

% ...



\section{Supplementary Information}
This section provides additional analyses, and explanations to support the main findings of the manuscript. It is organized into the following subsections to facilitate navigation and comprehension.

\section{Analytical calculation of CD coefficient from the first order nested commutator}
The first order CD coefficient, $\alpha_t$ is given by
\label{0.A}
\begin{widetext}
\begin{align}
\alpha_1 = -\frac{1}{4} \frac{ \sum_i h_i^2+ \sum_{i<j} J_{i j}^2 }{(1-\lambda)^2\left(\sum_i h_i^2+4 \sum_{i \neq j} J_{i j}^2\right) + \lambda^2\left[\sum_i h_i^4+\sum_{i \neq j} J_{i j}^4+6 \sum_{i \neq j} h_i^2 J_{i j}^2 +6 \sum_{i<j<k}\left(J_{i j}^2 J_{i k}^2+J_{i j}^2 J_{j k}^2+J_{i k}^2 J_{j k}^2\right)\right] } \, .
\end{align}
\end{widetext}
The detailed calculation of $\alpha_t$ is as follows. In the nested commutator expansion, the approximated CD term is given by 
\begin{equation}\label{eq:commutator}
    H_{cd}^{(l)}(\lambda) = i \dot{\lambda} \sum_{k=1}^l \alpha_k(\lambda) \underbrace{[ H_\text{ad}, [ H_\text{ad},\dots [ H_\text{ad},}_{2k-1} \partial_\lambda H_\text{ad} ] ] ],
\end{equation}
where $\ell$ is the order of the expansion and $\alpha_k$ are expansion coefficients to be found. The first order CD coefficient can be computed exactly for any Hamiltonian, provided the first two nested commutators $\mathcal{O}_1 = [H_{ad}, \partial_\lambda H_{ad}]$ and $\mathcal{O}_2 = [H_{ad}, \mathcal{O}_1]$, since $\alpha_1=-\Gamma_1 / \Gamma_2$, where $\Gamma_k=\left\|\mathcal{O}_k\right\|^2$ such that the CD term takes the form
\begin{equation}
    H_{C D}=i \dot{\lambda} \alpha_1 \mathcal{O}_1.
\end{equation}
For the case of the Hamiltonian used in Eq. (\ref{eq:adiabatic_hamiltonian}),
\begin{equation}
    \begin{aligned}
\mathcal{O}_1 & =\left[\sum_i \sigma^x_i, \sum_k h_k \sigma^z_k+\sum_{k l} J_{k<l} \sigma^z_k \sigma^z_k\right] \\
& = -2 i \sum_i h_i \sigma^y_i - 2 i \sum_{i<j} J_{i j}\left(\sigma^y_i \sigma^z_j+\sigma^z_i \sigma^y_j\right).
\end{aligned}
\end{equation}
Therefore, $\Gamma_1$ is given by\\

\begin{equation}
   \Gamma_1=4 \sum_i h_i^2+4 \sum_{i<j} J_{i j}^2 \, .
\end{equation}
Now, we need to calculate the denominator of the equation for $\alpha_1$, which follows from  $\mathcal{O}_2=(1-\lambda)\left[\sum_i \sigma^x_i, \mathcal{O}_1\right]+\lambda\left[\sum_i h_i \sigma^z_i + \sum_{i<j} J_{ij} \sigma^z_i \sigma^z_j, \mathcal{O}_1\right]$. Then, we need to evaluate two different commutators.
The first one can be written as
\begin{equation}
    \left[\sum_i \sigma^x_i, \mathcal{O}_1\right] = 4 \sum_i h_i \sigma^z_i + 8 \sum_{i<j} J_{i j} \sigma^z_i \sigma^z_j-8 \sum_{i k j} J_{i j} \sigma^y_i \sigma^y_j \, ,
\end{equation}
whereas the second one is given by the following equation as shown below
\vspace{1pt}
\begin{widetext}
\begin{equation}
    \begin{aligned}
\left[\sum_i h_i \sigma^z_i + \sum_{i<j} J_{ij} \sigma^z_i \sigma^z_j, \mathcal{O}_1\right] & =-4 \sum_i\left(h_i^2+\sum_j J_{i j}^2\right) \sigma^x_i-8 \sum_{i \neq j} h_i J_{i j} \sigma^x_i \sigma^z_j \\
&-8 \sum_{i<j<k}\left(J_{i j} J_{i j} \sigma^x_i \sigma^z_j \sigma^z_k+J_{i j} J_{j k} \sigma^z_i \sigma^x_j \sigma^z_k+J_{i k} J_{j k} \sigma^z_i \sigma^z_j \sigma^x_k\right).
\end{aligned}
\end{equation}
It follows that,
\begin{equation}
\begin{aligned}
\Gamma_2= & 16(1-\lambda)^2\left(\sum_i h_i^2+4 \sum_{i \neq j} J_{i j}^2\right)+16 \lambda^2\left(\sum_i h_i^4+\sum_{i \neq j} J_{i j}^4+6 \sum_{i \neq j}^1 h_i^2 J_{i j}^2 +6 \sum_{i<j<k}\left(J_{i j}^2 J_{i k}^2+J_{i j}^2 J_{j k}^2+J_{i k}^2 J_{j k}^2\right)\right).
\end{aligned}
\end{equation}
Hence, the CD term is given by
\begin{equation}
    H_{cd}^{(1)}(\lambda)=
2 \dot{\lambda} \alpha_1(\lambda)\left(\sum_i h_i \sigma^y_i+\sum_{i<j} J_{i j}\left(\sigma^y_i \sigma^z_j+\sigma^z_i \sigma^y_j\right)\right),
\end{equation}
where 
\begin{equation}
\alpha_1=-\frac{1}{4} \frac{\sum_i h_i^2+\sum_{i<j} J_{i j}^2}{R(t)},
\end{equation}
and
\begin{equation}
\begin{aligned}
    R(t)=&(1-\lambda)^2\left(\sum_i h_i^2+4 \sum_{i \neq j} J_{i j}^2\right)+\lambda^2\left(\sum_i h_i^4+\sum_{i \neq j} J_{i j}^4+ 6 \sum_{i \neq j}^1 h_i^2 J_{i j}^2 +6 \sum_{i<j<k}\left(J_{i j}^2 J_{i k}^2+J_{i j}^2 J_{j k}^2+J_{i k}^2 J_{j k}^2\right)\right).
\end{aligned}
\end{equation}
\end{widetext}

\subsection{Realizing $n-local$ terms using GMS gates}
\label{1.A}
Here, we show how to obtain $n-local$ terms using GMS gates and local rotations. For the GMS gate, Eq. (\ref{Eq.5}), let $\phi =0$, then we have $U_{MS}^k(\theta, 0) = e^{  {- i \theta} S_x^2}$ where $S_x = \sum_{j,k} \sigma^{x}$ which produces the $2-local$ terms between qubits $j-k$. With some algebra, it can be shown that a local operation applied to $l-$th qubit supplemented by GMS gate can produce cross Pauli terms as
\begin{align}
U_{j,k}(\theta) = e^{ \left[ {- i \theta} S_x^2 \right]} {\sigma_l^z} e^{ \left[ { i \theta} S_x^2 \right]}.
\end{align}
For $j=l$, we have
\begin{equation}
 U_{j,k}(\theta) {\sigma_l^z} U_{j,k}^{\dagger}(\theta) = [\sigma_l^z\cos(\theta/2) - \sigma_l^y\sigma_k^x \sin(\theta/2)]\delta_{j,l}.
\end{equation}
For a generalization to obtain $n-local$ terms, we can make use of the pattern of $\sigma_l^{z(y)}$ terms appearing for odd or even iterations given in Table. \ref{Tab.1}:
If we apply the transformation $n$ times, the $n$th iteration will give us
\begin{widetext}
\begin{align}
I^{(n)}_l (\theta) &=\prod_{i_1, i_2} U_{i_1,i_2}(\theta) {\sigma_l^z} U_{i_1,i_2}^{\dagger}(\theta) = \sigma_l^z\cos^{n-1}(\theta/2)  - \sum_{l} \sigma_{l}^y\sigma_{i_2}^x \cos^{n-2}(\theta/2) \sin(\theta/2) \nonumber \\& \quad - \sum_{l \neq i_2} \sigma_{l}^z\sigma_{i_2}^x\sigma_{i_3}^x \cos^{n-3}(\theta/2)\sin^2(\theta/2)  + \sum_{l \neq i_2 \neq i_3} \sigma_{l}^y\sigma_{i_2}^x\sigma_{i_3}^x\sigma_{i_4}^x \cos^{n-4}(\theta/2)\sin^3(\theta/2) \nonumber  \\
&\quad \vdots \nonumber \\
& \quad + \mathbf{1_{\text{even}}}_{(n)} \sum_{i_{(j)}} (-1)^{j/2} \sigma_{l}^y \sigma_{i_2}^x \dots \sigma_{i_{n-1}}^x \cos^{(n-j)}(\theta/2)\sin^{(j-1)}(\theta/2) \nonumber \\
&\quad + \mathbf{1_{\text{odd}}}_{(n)} \sum_{i_{(j)}} (-1)^{(j-1)/2} \sigma_{l}^z \sigma_{i_2}^x\sigma_{i_3}^x \dots \sigma_{i_n}^x \cos^{(n-j)}(\theta/2)\sin^{(j-1)}(\theta/2) \nonumber \\
&\quad \vdots \nonumber \\
&\quad (-)^{\lfloor n/2 \rfloor} \sum_{\substack{l \neq i_2, i_2 \neq i_3 \neq ... \neq i_n}} \sigma_{l}^y \sigma_{i_2}^x ... \sigma_{i_n}^x \sin^{n-1}(\theta/2).
\end{align}
\end{widetext}
where $\lfloor n/2 \rfloor = m$ and, $m\leq n < m+1$. 
\subsection{General building blocks for $n-local$ terms}
\label{1.B}
\begin{table}
\begin{tabular}{|c c| cccc  c| cccc  c| cccc | cccc c|  cccc c|  cccc c| cccc c| cccc c| cccc c| cccc c|}
 \hline
Even  &&&&& Odd &&&&& $\sigma_l$ type &&&&& Sign    \\ [0.5ex] 
 \hline
------- & & & &&   1 &&&&& $\sigma^z$ &&&&&   +   \\ 
 \hline
2 & & & &&   ------- &&&&& $\sigma^y$ &&&&&   -   \\ 
 \hline
------- & & & &&   3 &&&&&  $\sigma^z$ &&&&&   -   \\ 
 \hline
4 & & & &&   ------- &&&&& $\sigma^y$ &&&&&   +   \\ 
 \hline
------- & & & &&  5 &&&&& $\sigma^z$ &&&&&   +   \\ 
 \hline
6 & & & &&   ------- &&&&& $\sigma^y$ &&&&&   -  \\ 
 \hline
------- & & & &&   7 &&&&& $\sigma^z$ &&&&&   -   \\ 
 \hline
8 & & & &&   ------- &&&&& $\sigma^y$ &&&&&   +   \\ 
 \hline
------- & & & &&   9 &&&&& $\sigma^z$ &&&&&   +  \\ 
 \hline
10 & & & &&   ------- &&&&& $\sigma^y$ &&&&&   -  \\ 
 \hline
\end{tabular}
\caption{Patterns originating from even and odd local terms.}
\label{Tab.1}
\end{table}
\begin{figure*}
\centering
\includegraphics[width=1\textwidth]{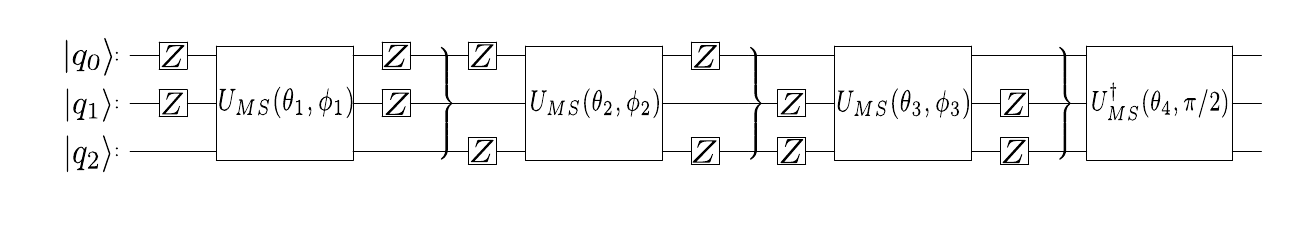}
\caption{Quantum circuit to realize in-homogenous $2-local$ terms in a simple example of a 3-qubit Hamiltonian, Eq. \ref{Eq.6} utilizing 3-qubit MS gates as analog blocks and Pauli Z gates as local rotations.}
\label{fig:1.D}
\end{figure*}
We can generate different types of $n-local$ terms by performing a basis transformation.  We can define a new basis $\hat{u}_j = \hat{i} \cos \phi  + \hat{j} \sin \phi $, so that
\begin{equation}
\sum_j \sigma_{j}^u = \sum_j\sigma_j^x\cos \phi + \sum_j\sigma_j^y\sin \phi.
\end{equation}
Substituting in Eq. (\ref{Eq.5}) we obtain
\begin{equation}
U_{MS}^n(\theta, \phi) = \exp \left[ -\frac{i \theta}{4} (S_{j})^2 \right],
\end{equation}
where $S_{j} = \sum_j^n  \sigma_{j}^u$. Now, we can express the gate as
\begin{equation}
U_{j,k} = \prod_{j,k}\exp \left[ -\frac{i \theta}{4}  \sigma_{j}^u \sigma_{k}^u \right],
\end{equation}
The generalized operation of this GMS gate to any pauli operator $\sigma_j^{\alpha}$ can be given as
\begin{equation}
I(\theta) = \prod_{j,k} e^{\left[ -{i \theta}  \sigma_{j}^u \sigma_{k}^u \right]} \sigma_j^{\alpha} e^{ \left[ {i \theta}  \sigma_{j}^u \sigma_{k}^u \right]},
\end{equation}
where $\alpha = x,y,z$.
Following the method in the last section and using the following commutation relations 
\begin{align}
[\sigma_j^{x}, \sigma_{j}^u ] = 2i\sigma_j^z \sin \phi, \; \; \; \; [\sigma_j^{y}, \sigma_{j}^u ] = -2i\sigma_j^z \cos \phi,\nonumber \\
[\sigma_j^{z}, \sigma_{j}^u ] = 2i\sigma_j^{\perp},\; \; \; \; \text{where} \;\; \sigma_j^{\perp} = \sigma_j^y\cos \phi - \sigma_j^x\sin \phi,
\end{align}
and
\begin{align}
[\sigma_{j}^u, [\sigma_j^{x}, \sigma_{j}^u ] ] = -4\sigma_j^{\perp} \sin \phi, \; \; \; \; [\sigma_j^{y}, \sigma_{j}^u ] = 4\sigma_j^{\perp} \cos \phi,\nonumber \\
[\sigma_{j}^u, [\sigma_j^{z}, \sigma_{j}^u ] ] = -4\sigma_j^{z},\; \; \; \; \text{where} \;\; \sigma_j^{\perp} = \sigma_j^y\cos \phi - \sigma_j^x\sin \phi,
\end{align}
we can obtain the generalization to $n$ particles giving us
\begin{widetext}
\begin{align}
\prod_{j,k} e^{\left[ -{i \theta}  \sigma_{j}^u \sigma_{k}^u \right]} \sigma_j^{\perp} e^{ \left[ {i \theta}  \sigma_{j}^u \sigma_{k}^u \right]} = & \quad + \mathbf{1_{\text{even}}}_{(n)} \sum_{i_{(j)}} (-1)^{j/2} \sigma_{l}^z \sigma_{i_2}^u \dots \sigma_{i_{n-1}}^u \cos^{(n-j)}(\theta/2)\sin^{(j-1)}(\theta/2) \nonumber \\
&\quad + \mathbf{1_{\text{odd}}}_{(n)} \sum_{i_{(j)}} (-1)^{(j-1)/2} \sigma_{l}^{\perp} \sigma_{i_2}^x\sigma_{i_3}^u \dots \sigma_{i_n}^u \cos^{(n-j)}(\theta/2)\sin^{(j-1)}(\theta/2) \nonumber \\
\end{align}
\end{widetext}

\subsection{Mapping the target Hamiltonian coefficients to the angles of GMS gate}
\label{1.C}
We perform time-evolution of the full CD Hamiltonian Eq. (\ref{Eq.6}) using trotterization obtaining
\begin{align}
\exp\{H'(\lambda)t/n\} 
=\bigg\{ \exp\bigg({-i\theta^{x,x} \sum_{i < j =1}^N  \sigma_{i}^x \sigma_j^x  }\bigg)\nonumber \\
 \cdot \exp\bigg({-i \theta^{x} \sum_{i=1}^N \sigma_i^x }\bigg) \cdot \exp\bigg({-i \theta^{z} \sum_{i=1}^N \sigma_i^z }\bigg) \nonumber \\ 
 \cdot \exp\bigg({-i\theta^{x,y}\sum_{i,j}^N\sigma_y^i \sigma_x^j}\bigg) \nonumber \\
 \cdot \exp\bigg({-i \theta^{x,y} \sum_{i < j=1}^N  \sigma_x^i \sigma_y^j }\bigg)\cdot \exp\bigg({-i \theta^{y} \sum_{i=1}^N \sigma_y^i }\bigg)
\bigg\}^n
\label{Eq.7}
\end{align}
where $\theta^{x,x}  = \lambda(t)  J t/n$, $\theta^{x} = \lambda(t) h t/n$, $\theta^z = (1-\lambda(t)) t/n$, $\theta^{x,y} =\theta^{y,x} = 2 J \dot{\lambda} {\alpha}_1(t) t/n$, $\theta^{y} = 2 \dot{\lambda} {\alpha}_1(t) ht/n$, where $t$ is the evolution time and $n$ is the number of trotter steps. To map these Hamiltonian coefficients to the angles of the GMS gate we expand the GMS gate Hamiltonian Eq. (\ref{Eq.5}) obtaining $\theta \cos^2(\phi) /2 = \theta^{x,x}$ and $\theta \cos(\phi) \sin(\phi)/4 = \theta^{x,y}$ using which we can realize the $2-local$ terms.

\subsection{Inhomogeneous case}
\label{1.D}

Inhomogeneous QUBO Hamiltonians can be generated by the Homogeneous analog blocks as explained in the main text. A simple example of realizing $2-local$ terms of a 3 qubit QUBO problem is shown in Fig. \ref{fig:1.D}. To introduce inhomogeneity across any $k$-qubit interactions within a $N$-qubit system, it is necessary to employ $k(k-1)/2$ analog blocks \cite{Adrian.2020} along with local rotations. Therefore, in this example, we need 3 analog blocks. The unitary generated by the quantum circuit is
\begin{align}
U(t_i) \approx  \; exp\bigg[\frac{i}{2} \bigg( t_1 \sigma_1^x\sigma_2^x + t_2 \sigma_2^x\sigma_3^x + t_3 \sigma_1^x\sigma_3^x \nonumber\\
+ t_4 (\sigma_1^y\sigma_2^y + \sigma_2^y\sigma_3^y + \sigma_1^y\sigma_3^y) 
+ t_5 (\sigma_1^x\sigma_2^y + \sigma_2^x\sigma_1^y)\nonumber \\
+ t_6 (\sigma_1^x\sigma_3^y + \sigma_3^x\sigma_1^y)
+ t_7 (\sigma_2^x\sigma_3^y + \sigma_3^x\sigma_2^y)\bigg)\bigg] \nonumber \\
- t_8 (\sigma_1^y\sigma_2^y + \sigma_2^y\sigma_3^y + \sigma_1^y\sigma_3^y) \bigg)\bigg].
\end{align}
where

$t_1 = -\theta_1\cos^2\phi_1 + {\theta_2} \cos^2\phi_2 +{\theta_3} \cos^2\phi_3$,

$t_2 = -{\theta_1} \cos^2\phi_1 + {\theta_2} \cos^2\phi_2 - {\theta_3} \cos^2\phi_3$,

$t_3 = {\theta_1} \cos^2\phi_1 - {\theta_2} \cos^2\phi_2 + {\theta_3} \cos^2\phi_3 $,

$t_4 = -{\theta_1} \sin^2\phi_1 - {\theta_2} \sin^2\phi_2 - {\theta_3} \sin^2\phi_3$,

$t_5 = -{\theta_1} \cos\phi_1\sin\phi_2 + {\theta_2} \cos\phi_1\sin\phi_2 + {\theta_3} \cos\phi_1\sin\phi_2$,

$t_6 = {\theta_1} \cos\phi_1\sin\phi_3 - {\theta_2} \cos\phi_1\sin\phi_3 + {\theta_3} \cos\phi_1\sin\phi_3 $,

$t_7 = {\theta_1} \cos\phi_2\sin\phi_3 + {\theta_2} \cos\phi_2\sin\phi_3 - {\theta_3} \cos\phi_2\sin\phi_3 $,

$t_8 = t_4$.

As the problem size grows, an increasing number of analog blocks are required per ion set to induce non-homogeneity, which suggests that enlarging the block size may not effectively reduce circuit depth. To determine the optimal block size for maximum circuit compression, we conducted an analysis, illustrated in Fig. \ref{fig:4}. We discovered that the ideal size for the analog block in this specific problem is a 2-qubit GMS gate $U_{MS}^{2}(\theta, \phi)$. Expanding the block beyond four qubits tends to be counterproductive. Furthermore, if hardware allows for the generation of non-homogeneous interactions, enabling what are known as programmable quantum gates \cite{Ravid.2020, Peleg.2023}, then the scalability improvements for inhomogeneous problems could match those of homogeneous ones.

\newpage






\begin{thebibliography}{99}



\bibitem{Lucas.2014} L. Andrew, \textit{Ising formulations of many NP problems}, \href{https://www.frontiersin.org/articles/10.3389/fphy.2014.00005}{Frontiers in Physics, \textbf{2}, (2014)}.

\bibitem{Amin.2023} A. D. King et al., \textit{Quantum critical dynamics in a 5,000-qubit programmable spin glass}, \href{https://www.nature.com/articles/s41586-023-05867-2}{Nature \textbf{617}, 61--66 (2023)}.

\bibitem{Henriet.2020} L Henriet, L. Beguin, A. Signoles, T. Lahaye, A. Browaeys G.O. Reymond, and C. Jurczak, \textit{Quantum computing with neutral atoms}, \href{https://quantum-journal.org/papers/q-2020-09-21-327/}{ 	Quantum \textbf{4}, 327 (2020).}

\bibitem{Gomez2023arXiv} A. Gomez Cadavid, I. Montalban, A. Dalal, E. Solano, and N. N. Hegade, \textit{Efficient DCQO Algorithm within the Impulse Regime for Portfolio Optimization}, \href{https://arxiv.org/abs/2308.15475}{arXiv:2308.15475 \textbf{[quant-ph]} (2023).}

\bibitem{Guan2023arXiv} H. Guan, F. Zhou, F. Albarrán-Arriagada, X. Chen, E. Solano, N. N Hegade, H.-L. Huang, \textit{Single-Layer Digitized-Counterdiabatic Quantum Optimization for $p$-spin Models}, \href{https://arxiv.org/abs/2311.06682}{arXiv:2311.06682 \textbf{[quant-ph]} (2023).}

\bibitem{Moll.2018}  N. Moll et al., \textit{Quantum optimization using variational algorithms on near-term quantum devices}, \href{https://doi.org/10.1088/2058-9565/aab822}{Quantum Sci. Technol. \textbf{3}, 030503 (2018)}.

\bibitem{Wiebe.2012} N. Wiebe, D. Braun, and S. Lloyd, \textit{Quantum Algorithm for Data Fitting}, \href{https://doi.org/10.1103/PhysRevLett.109.050505}{Phys. Rev. Lett. \textbf{109}, 050505 (2012)}, arXiv:\href{https://arxiv.org/abs/1204.5242}{1204.5242}.

\bibitem{Montanaro.2016} A. Montanaro, \textit{Quantum algorithms: an overview}, \href{https://doi.org/10.1038/npjqi.2015.23}{npj Quantum Inf. \textbf{2}, 15023 (2016)}, arXiv:\href{https://arxiv.org/abs/1511.04206}{1511.04206}.

\bibitem{Jens.2024} Niklas Pirnay, Vincent Ulitzsch, Frederik Wilde,Jens Eisert, and Jean-Pierre Seifert, \textit{An in-principle super-polynomial quantum advantage for approximating combinatorial optimization problems via computational learning theory}, \href{https://www.science.org/doi/10.1126/sciadv.adj5170}{Sci. Adv. \textbf{10}, 11 (2024)}

\bibitem{Naren.2021} N. N Hegade, K. Paul, Y. Ding, M. Sanz, F. Albarrán-Arriagada, E. Solano and Xi Chen, \textit{Shortcuts to Adiabaticity in Digitized Adiabatic Quantum Computing}, \href{https://journals.aps.org/prapplied/abstract/10.1103/PhysRevApplied.15.024038}{Phys. Rev. Applied \textbf{15}, 024038 (2021)}.

\bibitem{Naren.2022} N. N. Hegade, X. Chen, and E. Solano, \textit{Digitized counterdiabatic quantum optimization}, \href{https://link.aps.org/doi/10.1103/PhysRevResearch.4.L042030}{Phys. Rev. Res. \textbf{4}, L042030 (2023).}

\bibitem{Pranav.2023} P. Chandarana, N. N. Hegade, I. Montalban, E. Solano, and X. Chen, \textit{Digitized Counterdiabatic Quantum Algorithm for Protein Folding}, \href{https://journals.aps.org/prapplied/abstract/10.1103/PhysRevApplied.20.014024}{Phys. Rev. Applied \textbf{20}, 014024 (2023).}

\bibitem{DCQF.2023} N. N. Hegade, and Enrique Solano, \textit{Digitized-counterdiabatic quantum factorization}, \href{https://arxiv.org/pdf/2301.11005.pdf}{arXiv:2301.11005v1 (2023).}

\bibitem{Adrian.2020} A. Parra-Rodriguez, P. Lougovski, L. Lamata, E. Solano, and M. Sanz, \textit{Digital-analog quantum computation}, \href{https://journals.aps.org/pra/abstract/10.1103/PhysRevA.101.022305}{Phys. Rev. A \textbf{101}, 022305 (2020)}.

\bibitem{Pagano.2021} Z. Davoudi, N. M. Linke, and G. Pagano, \textit{Toward simulating quantum field theories with controlled phonon-ion dynamics: A hybrid analog-digital approach}, \href{https://journals.aps.org/prresearch/abstract/10.1103/PhysRevResearch.3.043072}{Phys. Rev. Research \textbf{3}, 043072  (2021)}.

\bibitem{Solano.2016} L. Garc\' ia-\'Alvarez, U. Las Heras, A. Mezzacapo, M. Sanz, E. Solano, and L. Lamata, \textit{Quantum chemistry and charge transport in biomolecules with superconducting circuits}, \href{https://doi.org/10.1038/srep27836}{Sci. Rep. \textbf{6}, 27836 (2016)}.

\bibitem{Jing.2021} J. Yu, J. C. Retamal, M. Sanz, E. Solano, and F. Albarrán-Arriagada, \textit{Superconducting circuit architecture for digital-analog quantum computing}, \href{https://doi.org/10.1140/epjqt/s40507-022-00129-y}{EPJ Quantum Technol. \textbf{9}, 9 (2022)}.

\bibitem{Sanz.2023} M. Garcia-de-Andoin, Á. Saiz, P. Pérez-Fernández, L. Lamata, I. Oregi, and M. Sanz, \textit{Digital-Analog Quantum Computation with Arbitrary Two-Body Hamiltonians}, \href{https://arxiv.org/abs/2307.00966}{arXiv:2307.00966 (2023)}.

\bibitem{Ana.2020} A. Martin, L. Lamata, E. Solano, and M. Sanz, \textit{Digital-analog quantum algorithm for the quantum Fourier transform}, \href{https://journals.aps.org/prresearch/abstract/10.1103/PhysRevResearch.2.013012}{Phys. Rev. Research \textbf{2}, 013012 (2020)}.

\bibitem{Ana.2022} D. Headley, T. Müller, A. Martin, E. Solano, M. Sanz, and F. K. Wilhelm, \textit{Approximating the quantum approximate optimization algorithm with digital-analog interactions}, \href{https://journals.aps.org/pra/abstract/10.1103/PhysRevA.106.042446}{Phys. Rev. A \textbf{106}, 042446 (2022)}.

\bibitem{Ana.2023} A. Martin, R. Ibarrondo, and M. Sanz, \textit{Digital-Analog Co-Design of the Harrow-Hassidim-Lloyd Algorithm}, \href{https://journals.aps.org/prapplied/abstract/10.1103/PhysRevApplied.19.064056}{Phys. Rev. Applied \textbf{19}, 064056 (2023)}.

\bibitem{Tasio.2021} T. Gonzalez-Raya, R. Asensio-Perea, A. Martin, L. C. Céleri, M. Sanz, P. Lougovski, and E. F. Dumitrescu, \textit{Digital-Analog Quantum Simulations Using the Cross-Resonance Effect}, \href{https://journals.aps.org/prxquantum/abstract/10.1103/PRXQuantum.2.020328}{PRX Quantum \textbf{2}, 020328 (2021)}.

\bibitem{David.2014} D. Hayes, S. T Flammia and M. J Biercuk \textit{Programmable quantum simulation by dynamic Hamiltonian engineering}, \href{https://iopscience.iop.org/article/10.1088/1367-2630/16/8/083027}{New Journal of Physics \textbf{16}, 083027(2014)}.

\bibitem{Rajabi.2019} F. Rajabi, S. Motlakunta, Chung-You Shih, N. Kotibhaskar, Q. Quraishi, A. Ajoy, and R. Islam  \textit{Dynamical Hamiltonian engineering of 2D rectangular lattices in a one-dimensional ion chain}, \href{https://www.nature.com/articles/s41534-019-0147-x}{npj Quantum Information \textbf{5}, 32 (2019)}.

\bibitem{Kumar.2023} S. Kumar, N. N. Hegade, E. Solano, F. Albarrán-Arriagada, and G. A. Barrios, \textit{Digital-analog quantum computing of fermion-boson models in superconducting circuits}, \href{https://doi.org/10.48550/arXiv.2308.12040}{arXiv:2308.12040 \textbf{[quant-ph]} (2023).}

\bibitem{Elfing.2024} C. Chevallier, J. Vovrosh, J. de Hond, M. Dagrada, A. Dauphin, and V. E. Elfving, \textit{Variational protocols for emulating digital gates using analog control with always-on interactions}, \href{https://arxiv.org/abs/2402.07653}{arXiv:2402.07653 \textbf{[quant-ph]} (2024).}

\bibitem{Mølmer.1999} A. Sørensen and K. Mølmer, \textit{Quantum Computation with Ions in Thermal Motion}, \href{https://journals.aps.org/prl/abstract/10.1103/PhysRevLett.82.1971}{Phys. Rev. Lett. \textbf{82}, (1971).}

\bibitem{Sørensen.1999} K. Mølmer and A. Sørensen, \textit{Multiparticle Entanglement of Hot Trapped Ions}, \href{https://journals.aps.org/prl/abstract/10.1103/PhysRevLett.82.1835}{Phys. Rev. Lett. \textbf{82},  1835(1971).}

\bibitem{Monz.2011} T. Monz et al, \textit{14-Qubit Entanglement: Creation and Coherence}, \href{https://journals.aps.org/prl/abstract/10.1103/PhysRevLett.106.130506}{Phys. Rev. Lett. \textbf{106}, 130506 (2011).}

\bibitem{Nam.2018} D. Maslov and Y. Nam, \textit{Use of global interactions in efficient quantum circuit constructions}, \href{https://iopscience.iop.org/article/10.1088/1367-2630/aaa398}{New Journal of Physics \textbf{20}, 033018 (2018).}

\bibitem{Rice.2003} M. Demirplak, and S. A. Rice, \textit{Adiabatic Population Transfer with Control Fields}, \href{https://pubs.acs.org/doi/10.1021/jp030708a}{J. Phys. Chem. A \textbf{107}, 46, 9937–9945 (2003)}.

\bibitem{Berry.2009} M. V. Berry, \textit{Transitionless quantum driving}, \href{https://iopscience.iop.org/article/10.1088/1751-8113/42/36/365303/meta}{J. Phys. A: Math. Theor. \textbf{42} 365303 (2009)}.

\bibitem{Muga.2010} X. Chen, I. Lizuain, A. Ruschhaupt, D. Guéry-Odelin, and J. G. Muga, \textit{Shortcut to Adiabatic Passage in Two- and Three-Level Atoms}, \href{https://journals.aps.org/prl/abstract/10.1103/PhysRevLett.105.123003}{Phys. Rev. Lett. \textbf{105}, 123003 (2010)}.

\bibitem{Campo.2013} A. del Campo, \textit{Shortcuts to Adiabaticity by Counterdiabatic Driving}, \href{https://journals.aps.org/prl/abstract/10.1103/PhysRevLett.111.100502}{Phys. Rev. Lett. \textbf{111}, 100502 (2013)}.

\bibitem{Sels2017PNAS} D. Sels and A. Polkovnikov, \textit{Minimizing irreversible losses in quantum systems by local counterdiabatic driving}, \href{https://www.pnas.org/doi/10.1073/pnas.1619826114}{PNAS \textbf{114}, E3909 (2017).}

\bibitem{Claeys2019PhysRevLett} P. W. Claeys, M. Pandey, D. Sels, and A. Polkovnikov, \textit{Floquet-Engineering Counterdiabatic Protocols in Quantum Many-Body Systems}, \href{https://journals.aps.org/prl/abstract/10.1103/PhysRevLett.123.090602}{Phys. Rev. Lett. \textbf{123}, 090602 (2019).}

\bibitem{Xie2022PhysRevB} Q. Xie, K. Seki, and S. Yunoki,\textit{Variational counterdiabatic driving of the Hubbard model for ground-state preparation}, \href{https://journals.aps.org/prb/abstract/10.1103/PhysRevB.106.155153}{Phys. Rev. B \textbf{106}, 155153 (2022).}

\bibitem{Kim.2023} Z. Cai, C.Y. Luan, L. Ou, H. Tu, Z. Yin, J.N. Zhang and K. Kim, \textit{Entangling gates for trapped-ion quantum computation and quantum simulation}, \href{https://link.springer.com/article/10.1007/s40042-023-00772-3}{ J. Korean Phys. Soc. \textbf{82}, 882–900 (2023).}

\bibitem{Casanova2012PhysRevLett} J. Casanova, A. Mezzacapo, L. Lamata, and E. Solano, \textit{Quantum Simulation of Interacting Fermion Lattice Models in Trapped Ions}, \href{https://journals.aps.org/prl/abstract/10.1103/PhysRevLett.108.190502}{ Phys. Rev. Lett. \textbf{108}, 190502 (2012).}

\bibitem{IonQ.2024} \textit{IonQ: Algorithmic Qubits}, \href{https://ionq.com/resources/algorithmic-qubits-a-better-single-number-metric}{https://ionq.com/resources/algorithmic-qubits-a-better-single-number-metric}.

\bibitem{QiskitAer.2021} \textit{Qiskit Aer: Simulator Backends}, \href{https://qiskit.org/ecosystem/aer/apidocs/aer_noise.html}{Qiskit Aer API Documentation}.

\bibitem{Bermudez.2017}A. Bermudez, \etal, \textit{Assessing the Progress of Trapped-Ion Processors Towards Fault-Tolerant Quantum Computation}, \href{https://journals.aps.org/prx/abstract/10.1103/PhysRevX.7.041061}{Phys. Rev. X \textbf{7}, 041061 (2023)}.


\bibitem{Ming.2023} R. Blümel, A. Maksymov, and M. Li, \textit{Toward a Mølmer Sørensen Gate With .9999 Fidelity}, \href{https://arxiv.org/abs/2311.15958}{	arXiv:2311.15958 \textbf{[quant-ph]} (2023)}.

\bibitem{Peleg.2023} Y. Shapira, L. Peleg, D. Schwerdt, J. Nemirovsky, N. Akerman, A. Stern, A. B. Kish, and R. Ozeri, \textit{Fast design and scaling of multi-qubit gates in large-scale trapped-ion quantum computers} \href{https://arxiv.org/abs/2307.09566}{ arXiv:2307.09566 (2023).}

\bibitem{Nam.2020} N. Grzesiak et al, \textit{Efficient arbitrary simultaneously entangling gates on a trapped-ion quantum computer}, \href{https://www.nature.com/articles/s41467-020-16790-9}{ Nat Commun \textbf{11}, 2963 (2020).}

\bibitem{Nam.2022} N. Grzesiak1, A. Maksymov1, P. Niroula, and Y. Nam \textit{Efficient quantum programming using EASE gates on a trapped-ion quantum computer}, \href{https://quantum-journal.org/papers/q-2022-01-27-634/}{ Quantum \textbf{6}, 634 (2022).}

\bibitem{Ravid.2020} Y. Shapira, R. Shaniv, T. Manovitz, N. Akerman, L. Peleg, L. Gazit, R. Ozeri, and A. Stern. \textit{Theory of robust multiqubit nonadiabatic gates for trapped ions}, \href{https://journals.aps.org/pra/abstract/10.1103/PhysRevA.101.032330}{Phys. Rev. A 101, 032330}

\bibitem{Shapira.2020} Tom Manovitz, Yotam Shapira, Nitzan Akerman, Ady Stern, and Roee Ozeri, \textit{Quantum Simulations with Complex Geometries and Synthetic Gauge Fields in a Trapped Ion Chain}, \href{https://link.aps.org/doi/10.1103/PRXQuantum.1.020303}{PRX Quantum \textbf{1}, 020303 (2020).}





\end{thebibliography}
\end{document}